\documentclass[apjl]{emulateapj}

\shorttitle{Variability of X - ray flares}
\shortauthors{Sonbas et al.}

\begin{document}


\title{A New Correlation between GRB X-Ray Flares and the Prompt Emission}


\author{E. Sonbas\altaffilmark{1,2}}
\affil{$^1$University of Adiyaman, Department of Physics, 02040 Adiyaman, Turkey}
\affil{$^2$NASA Goddard Space Flight Center, Greenbelt, MD 20771, USA}  

\author{G. A. MacLachlan\altaffilmark{3}, A. Shenoy\altaffilmark{3}, K.S. Dhuga\altaffilmark{3} and W. C. Parke\altaffilmark{3}}
\affil{$^3$Department of Physics, The George Washington University, Washington, DC 20052, USA}
\email{edasonbas@yahoo.com}

\begin{abstract}
From a sample of GRBs detected by the $Fermi$ and $Swift$ missions, we have extracted the minimum variability time scales for temporal structures in
the light curves associated with the prompt emission and X-ray flares. A comparison of this variability time scale with pulse parameters such as rise times, 
determined via pulse-fitting procedures, and spectral lags, extracted via the cross-correlation function (CCF), indicate a tight correlation between these 
temporal features for both the X-ray flares and the prompt emission. These correlations suggests a common origin for the production of X-ray flares and the 
prompt emission in GRBs. 
\end{abstract}

\keywords{X-ray flares, Gamma-ray bursts}

\section{Introduction}

The occurrence of X-ray flares (XRFs), associated with a large percentage of the GRBs detected by $Swift$ is now well established (Burrows et al. 2005;
Romano et al. 2006; Falcone et al. 2006; Chincarini et al. 2007). Interest now concentrates on how this flaring is related to the physics of the 
prompt emission, early afterglow, the transition between these phases via internal (and possible external) shock activities of GRBs, and the 
variability of the central engine itself (Rees \& Meszaros 1994; Kobayashi et al. 1997; Panaitescu et al. 1999; Zhang et al. 2006; 
Maxham \& Zhang 2009; Yu \& Dai 2009). 

As has been noted in a number of studies (Burrows et al. 2005; Nousek et al. 2006; O'Brien et al. 2006; Willingale et al. 2007), the X-ray 
light curves observed by $Swift$/XRT follow a similar pattern; essentially comprised of a prompt exponential decay followed by a steep power-law decay over a 
certain time scale. For most GRBs, the steep decay is followed by a shallow plateau that gradually gives way to another decreasing phase during which the 
X-ray flux decays according to a different power-law over a time scale that is significantly longer compared to the prompt emission and the early afterglow. 
XRFs are known to occur predominantly during the steeply declining phase of the X-ray light curve but flares during the plateau portion of the light curve are not uncommon. 
Empirically, the behavior of the composite light curve is consistent with the presence of two emission processes that overlap in time (Willingale et al. 2007): a short-duration 
episode in addition to an episode longer in duration but lower in luminosity. However, the underlying mechanisms that produce the flaring activity are not fully understood. Typical 
questions that arise include (a) are XRFs related to the late activity of the central engine, and (b) is the same (internal) shock mechanism responsible for 
both the prompt emission and the flaring activity? These questions have been tackled in different ways but focus primarily on linking the observed temporal 
and spectral properties of prompt emission in long bursts to similar properties seen in bursts exhibiting XRFs: Examples of these properties and/or relations 
include extending the lag-luminosity relation to X-ray flares, comparing pulse-profiles of temporal structures in the prompt emission and X-ray flares, and 
studies of evolution of spectral lag and the comparison of spectral hardness of XRFs with that of the underlying afterglow.
 
The lag-luminosity relation for XRFs has been investigated by Margutti et al. 2010 and was found to be consistent with the existing relation for the 
prompt emission (Ukwatta et al. 2012; Norris et al. 2007) suggesting that XRFs may share common origins with prompt emission. A very similar 
study (Sultana et al. 2012) makes a connection between the prompt emission data and the late afterglow X-ray data and suggests that the lag-luminosity relation is 
valid over a time scale well beyond the early steep-declining phase of the X-ray light curve. Maxham \& Zhang 2009 present a summary of the salient properties of 
XRFs and also show, using an internal shell collision model, that the main time histories of XRFs can be explained by the late activity of the central engine. Another 
study that hints at a connection between the prompt emission and the X-ray afterglow is that of Kocevski et al. 2007 in which the authors examined the evolution of 
pulse widths of the flares and found that the correlation between the widths of the pulses and time is consistent with the effects of internal shocks at ever increasing 
collision radii. Other techniques that seem to hold promise include the study of E$_{peak}$ evolution (Sonbas et al. 2012), the investigation of the relations predicted 
by various curvature models (Liang et al. 2006, Shenoy et al. 2012), and the time variability of bursts.  

In a recent wavelet analysis (MacLachlan et al. 2013) of the gamma-ray prompt emission from a sizable sample of long and short GRBs detected by the 
$Fermi$/GBM satellite, it was shown that a variability (related to the minimum time scale that separates white noise from red noise) of a few milliseconds 
is quite common. Moreover, it was demonstrated that there is a direct link between the shortest pulse structures as determined by the minimum time scale 
and pulse-fit parameters such as rise times. This type of analysis is quite easily extended to a larger sample of (long-duration) $Swift$ bursts, where the time 
variability and pulse-fit parameters for both the prompt emission and XRFs can be extracted and compared. In this paper, we report on the results of 
such an analysis.

\section{Data and Methodology}

The prompt emission light curves for a sample of GBM bursts were taken directly from the work of MacLachlan et al. 2013, who used a 
technique based on wavelets to determine the minimum time scale (MTS) at which scaling processes dominate over random noise processes. The authors 
associate this time scale with a transition from red-noise processes to parts of the power spectrum dominated by white-noise or random noise components. 
Accordingly, the authors note that this time scale is the shortest resolvable variability time for physical processes intrinsic to the GRB. We have used 
the same technique to extract the time scaling characteristics of the XRFs for a small sample of $Swift$ bursts (see Table 1). For the extraction of 
X-ray light curves, we used the method developed by Evans et al. 2009 in WT (windows timing) mode. Using the software tools available directly from their 
website ($\texttt http://www.swift.ac.uk/user objects/$), we extracted X-ray-flare light curves with different time bins. By constructing log-scale diagrams 
(plot of log(variance) of the signal vs. inverse frequency in octaves) for the sample, we have determined the minimum time scale above which scaling processes dominate 
over random intrinsic noise processes. A typical example of a bright XRF light curve and the associated log-scale diagram is shown in Figure 1. Note that the white-noise region 
(plateau region of the log-scale diagram) intersects the red-noise region (scaling region) at around octave 3.5 which corresponds to approximately 6 seconds 
for the light curve in question. As noted by MacLachlan et al. 2013, the time scale for the transition from the scaling region to the plateau region provides 
a measure of the smallest time variation for physical processes intrinsic to the GRB. We associate this time scale with the variability of the burst. 

\begin{table}[ht] 
\caption{Minimum variability times for the XRFs in the sample} % title of Table 
\centering % used for centering table 
\begin{tabular}{cccc} % centered columns (4 columns) 
\hline\hline %inserts double horizontal lines 
GRB Name & $\tau$ [sec] & $\delta\tau{^-}$ [sec] & $\delta\tau{^+}$ [sec]\\ [0.5ex] % inserts table %heading 
\hline % inserts single horizontal line 
GRB 050502B & 10.35 & 1.89 &2.98 \\ % inserting body of the table
GRB 050713A & 1.43 & 0.53 & 2.11 \\
GRB 050730 & 10.93 & 1.81 &2.70 \\
GRB 050822 & 5.83 & 2.13 & 7.99 \\
GRB 051117A & 13.14 & 2.91 &5.23 \\
GRB 060111A & 6.25 & 1.17 & 1.87 \\
GRB 060124 &  3.02 & 1.17 &5.33 \\
GRB 060204B & 2.64  & 0.69 & 1.43 \\ 
GRB 060210  & 3.78 & 0.97 & 2.02 \\ 
GRB 060312  & 1.21 & 0.46  & 1.97 \\ 
GRB 060418 & 2.13 &0.78 & 2.94 \\ 
GRB 060526 & 2.87 & 0.93 & 2.69 \\
GRB 060607A & 3.35 &0.78 & 1.44 \\ 
GRB 060714 & 2.20 & 0.75 & 2.36 \\ 
GRB 060904A & 2.92 & 1.03 & 3.57 \\  
GRB 060904B & 3.02 & 1.21 & 6.20\\  
GRB 060929 & 9.59 & 1.67 & 2.56 \\
GRB 070520B & 5.94 & 1.09 & 1.73 \\
GRB 070704 & 6.89 & 1.41 & 2.40 \\ 
\hline %inserts single line 
\end{tabular} 
\label{table:nonlin} % is used to refer this table in the text 
\end{table}

\begin{figure}
\epsscale{1.20}
\plotone{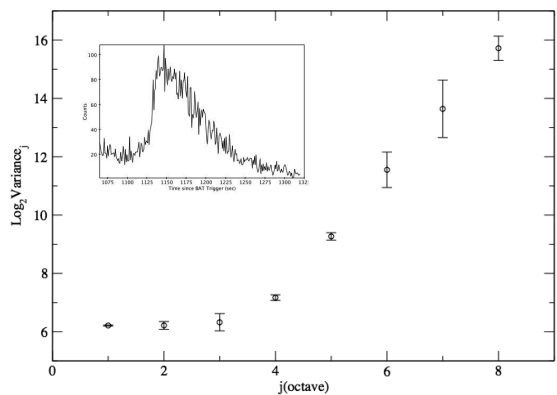}
\caption{Logscale diagram (and light curve; see inset) for the bright X-ray flare in GRB070520B: Log(Variance) of signal as a function of 
octave (inverse frequency). Plateau region is white noise and the sloped region is red noise. \label{fig1}}
%\vspace{0.3cm}
\end{figure}

%\begin{figure}
%\epsscale{1.00}
%\includegraphics[scale=0.29, angle = -90.0]{Fig2.ps}
%\caption{Logscale diagram for the bright X-ray flare in GRB070520B: Log(Variance) of signal as a function of 
%octave (frequency). Plateau region is white noise and the sloped region is red noise.\label{fig2}}
%\vspace{0.3cm}
%\end{figure}

%\vspace{0.6cm}

Using a particular functional form for pulse shapes, Margutti et al. 2010, have extracted a set of key pulse-fit parameters such as rise times, decay times, 
widths, and times since trigger for a set of bright XRFs detected by $Swift$/XRT. Their prime interest lay in the testing of (and extending) 
the validity of the lag-luminosity relation for XRFs. Our immediate interest in this study, however, focuses on their results for the various pulse-fit 
parameters such as rise times and pulse widths, because we can use these directly to compare with the variability time scales that we have extracted for the 
prompt emission and the X-ray flares. We note that the pulse rise times are invariably shorter than the pulse widths or decay times. To augment our sample 
we have also used the pulse-fit parameters from the work by Kocevski et al. 2007. The appropriate pulse-fit parameters for the prompt emission data were 
taken from the catalog produced by Bhat et al. 2012.

\section{Results and Discussion}

Following the work of MacLachlan et al. 2012, we have used a technique based on wavelets to extract a minimum time scale for a sample of GRBs detected by 
the $Fermi$ and $Swift$ missions. Shown in Figure 2 is a plot of the pulse rise-times versus the minimum time scale for the GRBs in our sample. We have 
plotted the data as observer-frame quantities because the redshift-dependent time dilation factor is the same for both variables: Black data points 
indicate the prompt emission data (with the pulse-fit parameters from Bhat et al. 2012); the blue and green points depict the XRF data with pulse-fit 
parameters taken from Kocevski et al. 2007 and Margutti et al. 2010) respectively. Also shown in the figure is a line depicting the equality of time 
scales. The best-fit line (not shown) leads to a slope of 1.26 $\pm$ 0.05.   The data show a strong correlation (Spearman correlation of 0.96 $\pm$ 0.02 
and a Kendall correlation of 0.79 $\pm$ 0.02) between pulse rise times and minimum time scales all the way from prompt emission to X-ray flares, i.e. more 
than three decades of variability time. This result extends the work of MacLachlan et al (2012), who examined prompt emission only, to the temporal 
domain covered by XRFs and reinforces their main conclusion that the two techniques, wavelets and pulse-fitting, can be used independently to extract a 
minimum time scale for physical processes of interest as long as close attention is paid to time binning and the proper identification of distinct pulses. In 
order to pursue the apparent connection between the temporal properties of prompt emission and the XRFs, we explore below the possible link between another 
temporal property, that of spectral lags, and the MTS.
   
%------------------------------------------------
\begin{figure}[htp]
%\fontsize{10}{12}\selectfont
%\begin{minipage}[b]{0.5\textwidth}
%\centering
%\vspace{-0.10in}
%\plotone[scale=0.29, angle = -90.0]{Trisevsmts.ps}
\includegraphics[scale=0.27, angle = 0.0]{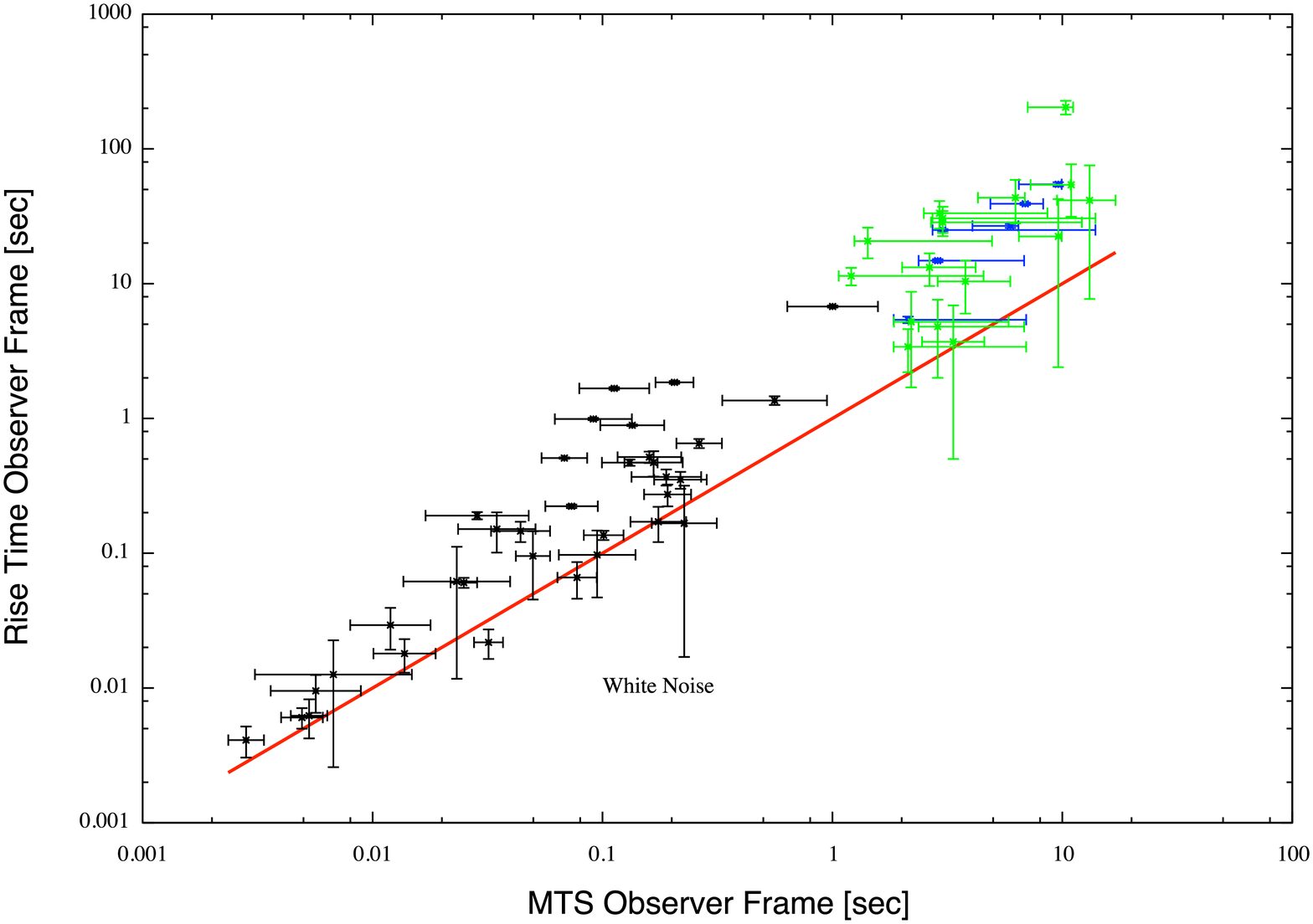}
%\caption{Rise time versus the minimum variability time scale in the observer frame with only smallest rise times included for Fermi/GBM data. The solid line indicates the equality of the respective scales.}}
\caption{Rise time versus the minimum variability time scale in the observer frame for a sample of GRBs: Black points (prompt emission); green and blue points (XRF data).
The solid line indicates the equality of the respective temporal scales.}
\label{fig2}
%\end{minipage}
\end{figure}

For the prompt emission data, we extracted spectral lags for various observer-frame energy bands using the CCF method described in detail by Ukwatta et al. (2012). Some 
of these results have been presented by Sonbas et al. 2012. Using the flare peak times reported by Margutti et al. (2010), we have also extracted the spectral lags for 
the XRFs between the energy bands 0.3-1 keV and the 3-10 keV respectively. A plot of the spectral lags versus the minimum time scale for the GRBs in our sample is 
shown in Figure 3. Black and magenta data points depict the prompt emission for long and short bursts; the blue points represent the XRF data. The red line indicates 
the best-fit (a slope of 1.44 $\pm$ 0.07) through the combined data set. The results clearly indicate a strong positive correlation (a Spearman correlation of 0.96 $\pm$ 0.05 
and a Kendall correlation of 0.86 $\pm$ 0.05) between the two temporal features, spectral lag and the MTS. Also shown in Figure 3 (see insert) is a plot of the pulse-rise times 
as function of the spectral lags. As expected, a positive correlation is observed but the scatter appears to be relatively large at the small time scales possibly indicating 
the difficulty in the identification and fitting of pulses at these scales. In addition, we note, as did MacLachlan et al 2012, that the uncertainties in the pulse rise times, 
quoted by Bhat et al 2012, are in many cases significantly smaller than the time binning of the lightcurves. We follow MacLachlan et al 2012 and adjust the uncertainties in the rise 
times by folding in quadrature the bin widths to the uncertainties given by Bhat et al 2012. With this minor adjustment, we argue that the observed correlations, taken as a whole, are 
suggestive of more than a trivial connection between the prompt emission and the XRFs. 

\begin{figure}[htp]
\fontsize{10}{10}\selectfont
\begin{minipage}[b]{0.5\textwidth}
\centering
\vspace{0.05in}
\includegraphics[scale=0.27, angle = 0.0]{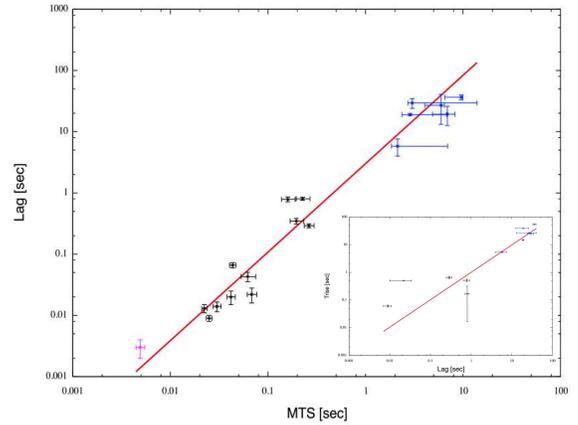}
%\label{fig3:a}
%\includegraphics[scale=0.27, angle = 0.0]{LagvsTrise.eps}
%\label{fig3:b}
\caption{Observer frame spectral lags and minimum variability time scales are plotted for prompt and flare emission: Black points (prompt emission for long 
bursts); magenta point prompt emission for short burst); and blue points (XRF data). (insert) Observer frame rise times as function of spectral lags for prompt 
and flare emission: Black points (prompt emission for long bursts); and blue points (XRF data). In both cases, the solid line indicates the best-fit to the data.}
\label{fig4}
\end{minipage}
\end{figure}

It is relatively straightforward to interpret the correlation between pulse parameters and the MTS in terms of the internal shock model in 
which the basic units of emission are assumed to be pulses that are produced via the collision of relativistic shells emitted by the central engine. Quilligan et al. 2002 in 
their study of the brightest BATSE bursts identified and fitted distinct pulses and showed a strong positive correlation between the number of pulses and the duration of the burst. 
More recent studies (Bhat \& Guiriec 2011; Hakkila \& Cumbee 2009; Hakkila \& Preece 2011) have provided further evidence for the pulse paradigm view of the prompt emission in GRBs. 
Maxham \& Zhang 2009, use the internal shell collision model to probe the spectral and temporal connection between the prompt emission and the XRFs. By assuming the Band function 
for the spectrum, an empirical temporal profile for the flares, and arbitrary central engine activity, they are able to explain the major temporal features of the XRFs, in particular, they 
note that the XRF time history reflects the time history of the central engine, which reactivates multiple times after the main prompt emission phase. Other authors 
(Narayan \& Kumar 2009) invoke relativistic outflow mechanisms to suggest that local turbulence amplified through Lorentz boosting leads to causally disconnected regions 
that in turn act as independent centers for the observed prompt emission. In more recent developments (Zhang \& Yan 2011, Vetere et al. 2006, Gao et al. 2012, MacLachlan et al. 2013) there is a suggestion that 
the variability may be composed of two distinct time scales; a rapidly varying component (order of milliseconds) embedded on a slower component (order of seconds) with the implication 
that these two components probe distinctly different aspects of GRB production and propagation. Similarly, simulation studies (Morsony et al. 2010) of the propagation of a 
jet through stellar material indicate that the temporal variability at different time scales is possibly related to the central engine and the propagation of the 
jet itself, and is measurable from the prompt emission. In the model reported by Zhang and Yan (2011), the authors invoke a magnetically dominated relativistic 
outflow to suggest that it is the slow component of the variability that is linked to the activity of the central engine and that the more rapidly varying component 
is associated with magnetic turbulence. While we are not in a position to distinguish between the aforementioned models, which incidentally are typically used to 
describe the variability only in the prompt emission, it is intriguing nonetheless that the observed correlation particularly that between the spectral lag and 
the MTS connects both the prompt emission and the flaring activity. 

Kocevski et al. (2007) suggest that the rise time of the X-ray flare pulse is related to the shell thickness as two shells 
collide after the second (faster) shell catches up with the (slower) first shell. The observed rise-time is estimated by:
%\begin{equation}
$\Delta t _r \approx \frac{\delta R}{2c \Gamma^{2}_{m}}$, 
%\end{equation}
where $\delta R$ is the thickness, $\Gamma_{m}$  is the relative Lorentz factor of the merged shells. Higher-latitude emission (for viewing angles less than the opening 
angle, $\theta$, of a conical jet) will be detected as broader pulses than lower-latitude ones. Following Zhang et al 2006, one can determine the decay time scale as the 
difference in light-travel time between photons emitted along the line of sight and the photons emitted at an angle along a shell of a given radius, R.
\begin{equation}
\Delta t_{decay}  \approx ( R / c ) (\theta^{2}/ 2) 
\end{equation}
For simplicity, we have omitted the redshift-dependent dilation factor. If we can assume the decay time scale is the spectral lag due to curvature, then the above arguments 
suggest a correlation between the lag and some measure of the variability which we associate with the MTS. While our interpretation is obviously speculative, the existence of the 
strong correlation, which we contend to be of astrophysical significance, warrants detailed theoretical investigation.  

As far as the extraction of the time variability directly from data is concerned, the wavelet method of MacLachlan et al. 2013 does not assume any temporal profile nor does it rely on identifying distinct pulses but instead uses the multi-resolution capacity of the wavelet technique to resolve the smallest significant temporal scale present in the light curves (of prompt emission and XRFs). These authors showed that the shortest pulse structures and the MTS track each other very closely for the prompt emission. In this work we have demonstrated that the spectral lag too tracks the MTS. Moreover, we have extended the work of MacLachlan et al 2012 to include both the prompt emission and the XRFs. This result, depicted in Figure 3 (supported by the data in Figure 2), provides new and compelling evidence that, as far as these temporal measures are concerned, the XRFs appear to be simple 'temporal extensions' of the pulse 
structures observed in the prompt emission.  

\section{Conclusions}

For a sample of long-duration GRBs detected by the $Fermi$/GBM and $Swift$ missions, we have extracted the minimum variability time scales and spectral lags for both 
prompt emission and XRF light curves. In addition, we have utilized the pulse-fit parameters presented by Margutti et al. 2010 and Kocevski et al. 2007 from their respective 
studies of XRFs. We compare the minimum variability time scale, extracted through a technique based on wavelets, both with the pulse rise times extracted through a 
fitting procedure, and spectral lags extracted via the CCF method. With these combined results, we have studied the relationship between key parameters that describe the temporal properties of a sample of prompt-emission and XRF light curves. Our main results are summarized as follows;
(1) The prompt emission and the XRFs exhibit a significant positive correlation between pulse rise times and the minimum time scale, with time scales ranging from several milliseconds to a few seconds respectively,
(2) The short-time variabilities in the prompt emission scale over time into the short-time variabilities in XRFs and 
(3) The spectral lag for both the prompt emission and the XRFs shows a strong positive correlation with the minimum variability time scale.   
Taken together these results are highly suggestive of a direct link between the mechanisms that lead to the production of XRFs and prompt emission in GRBs. 
\section{Acknowledgements}
This work made use of data supplied by the UK $Swift$ Science Data Centre at the University of Leicester. The work of ES was partially supported through the $Swift$ Mission 
(PI: N. Gehrels) and is gratefully acknowledged.

\end{document}